\documentclass[
 aip,
 cha,
 amsmath,amssymb,
 reprint
]{revtex4-1}

\usepackage{graphicx}
\usepackage{dcolumn}

\usepackage[utf8]{inputenc}
\usepackage[T1]{fontenc}
\usepackage{mathptmx}
\usepackage{etoolbox}

\usepackage[dvipsnames]{xcolor}

\makeatletter
\def\@email#1#2{
 \endgroup
 \patchcmd{\titleblock@produce}
  {\frontmatter@RRAPformat}
  {\frontmatter@RRAPformat{\produce@RRAP{*#1\href{mailto:#2}{#2}}}\frontmatter@RRAPformat}
  {}{}
}
\makeatother
\begin{document}

\title{A Panoramic Network Overview of Shortcut Impact on Power Grids}

\author{Yunju Choi}
\affiliation{Department of Energy Engineering, Korea Institute of Energy Technology (KENTECH), Naju, Jeonnam 58330, Republic of Korea}

\author{Daekyung Lee}
\affiliation{Supply Chain Intelligence Institute Austria, Metternichgasse 8, 1030 Vienna, Austria}

\author{Heetae Kim}
 \email[Corresponding author: ]{hkim@kentech.ac.kr}
\affiliation{Department of Energy Engineering, Korea Institute of Energy Technology (KENTECH), Naju, Jeonnam 58330, Republic of Korea}

\date{\today}

\begin{abstract}
    Synchronization of coupled oscillators on complex networks underlies critical functions in natural and engineered systems, including power-grid operation. While synchronization stability on fixed network topologies has been studied extensively, less is known about how an added transmission link affects stability depending on where it is placed relative to generation and consumption. 
We address this question using simplified ring-network power-grid models in which producer and consumer nodes are spatially organized into clusters. 
Long-range transmission links, which we refer to as shortcuts, are added between non-adjacent nodes, and synchronization stability is quantified by basin stability in second-order coupled-oscillator dynamics. 
We compare shortcut placements across four scenarios varying network size, power allocation, power-injection heterogeneity, and producer--consumer arrangement. 
When producers and consumers form two spatially separated clusters, center-to-center shortcuts do not reduce the mean basin stability in the configurations examined and increase it by up to 75\% relative to the shortcut-less baseline. 
Peripheral shortcuts instead reduce mean basin stability by up to 28\% in some configurations despite the added connection, producing Braess-like degradation.
In alternating multi-block networks, however, no single shortcut placement provides the same substantial enhancement, indicating that the center-to-center advantage does not directly generalize to more complex spatial arrangements.
Representative time-domain simulations further show faster decay of frequency deviations for center-to-center than for peripheral shortcuts. 
These results demonstrate that the effect of added connectivity depends critically on where the new link is placed relative to the spatial organization of generation and consumption.
\end{abstract}


\maketitle

\begin{quotation}
Adding a connection to a network can change its collective dynamics in ways that depend strongly on where the new link is placed. 
We study this effect in oscillator models of power grids, where stable operation requires the network to recover synchronization after disturbances. 
When producers and consumers form spatially separated groups, links connecting their centers are consistently favorable, while peripheral links can even reduce stability despite the additional connectivity. 
For more interleaved producer--consumer arrangements, however, this simple placement advantage weakens, showing that the effect of structural reinforcement depends on both link placement and the spatial organization of the network.
\end{quotation}

\section{Introduction}\label{sec:introduction}

From the firing of neurons to the rotation of power generators, the collective behavior of coupled oscillators emerges from the dynamics of individual units and is shaped by the structure of the network that connects them\cite{arenas2008synchronization, strogatz2000kuramoto, rodrigues2016kuramoto}.
How a given structure sustains or undermines synchronization has therefore become a central problem in network science, addressed in neural systems~\cite{varela2001brainweb, buzsaki2004neuronal, breakspear2010generative}, biological rhythms~\cite{yamaguchi2003synchronization, winfree1967biological, gonze2005spontaneous}, chemical oscillations~\cite{kiss2002emerging, taylor2009dynamical}, and large-scale infrastructure~\cite{motter2013spontaneous, albert2004structural, buldyrev2010catastrophic}.
Among these systems, power grids stand out because their synchronization carries a direct operational consequence.
The rotating generators of an alternating current (AC) power grid must remain synchronized at the rated frequency, which is the condition under which electricity is supplied reliably.
Their rotational dynamics can be described by second-order coupled-oscillator equations~\cite{filatrella2008analysis, dorfler2012synchronization}, so that keeping a grid synchronized is also a question about the structure of the network on which those oscillators are coupled.

This question has become more pressing as modern power systems undergo rapid structural transformation.
Electricity demand continues to grow and concentrates in urban and industrial areas, where shared infrastructure serves many consumers at once. 
Generation capacity expands in response, but the siting of large-scale power plants is constrained by geography, resource availability, and land use, so that plants are frequently built far from the areas they supply~\cite{cohon1980application, hoffacker2020local}.
This can produce spatially separated clusters of generation and consumption.
Transfer between these clusters can then concentrate on a few paths. 
Within the stable operating range, the magnitude of the phase angle difference across each line on such a path grows with the power the line carries relative to its coupling strength. 
Since the restoring stiffness of a line falls as that magnitude grows~\cite{dorfler2012synchronization, kundur2007power}, a configuration that loads a few lines heavily can be less able to return to synchrony after a disturbance.

A second transformation is changing how generation itself is distributed.
Many future-grid architectures involve a more spatially distributed allocation of generation capacity, in which a given amount of generation is supplied by a larger number of smaller units~\cite{rohden2012self, witthaut2022collective}.
Synchrony must then be maintained across more nodes, each assigned a smaller share of the total power.
How this shift affects stability is not settled by the number of sources alone, since the same aggregate generation can be distributed over a network in many different ways.

Assessing stability under these conditions has traditionally relied on component-based methods such as N-1 or N-2 contingency assessments, which evaluate the impact of specific line or generator outages~\cite{tpl2015001, kundur2007power, wood2013power}.
These methods evaluate how the system responds to an enumerated set of component failures, and they are effective for operational planning.
Basin stability addresses a complementary question, namely the probability that the network returns to synchrony after a finite perturbation drawn from a prescribed distribution of states~\cite{menck2013basin}.
Answering that question calls for a perspective that treats the coupled dynamics and the network topology of the grid together.

Representing a power grid as a graph meets this requirement by bringing the analytical tools of nonlinear dynamics to bear on the system as a whole, while the physical content of the grid is preserved in the parameters attached to that graph.
In this representation, generators and loads are nodes with a net power injection, and transmission lines are edges with a coupling strength.
Researchers have worked to make this representation quantitative, developing reduced models that retain the dynamics governing synchronization while remaining tractable at network scale.
A comparative analysis of existing power-grid models clarified how modeling assumptions shift predicted stability boundaries and under which conditions such reduced models reproduce grid behavior~\cite{nishikawa2015comparative}.
Stability has also been resolved down to individual nodes.
A single-node basin stability gives the probability that a network returns to synchrony after a finite perturbation applied to one node~\cite{menck2014dead}.
Applying it to grid models has shown that dead-end nodes and community structure each shape the stability of the nodes involved~\cite{menck2014dead, kim2015community}.

These results describe the stability of a grid whose structure is given. 
Changing that structure raises a separate question, because the effect of a change does not stay where the change is made. 
Studies of cascading failures have shown that removing or adding a single line redirects flows and can overload components far from the modified link \cite{witthaut2016critical, witthaut2013nonlocal}, and transient dynamics amplify such nonlocal effects beyond what static analyses predict \cite{simonsen2008transient}. 
The consequence of a structural change therefore depends on the network in which the change is embedded, not on the changed element alone.

One structural intervention that can affect synchronization stability is the addition of new transmission connections, which provide alternative paths for power flow and may relieve the loading of existing paths. 
In practice, however, public opposition, permitting difficulties, environmental constraints, and investment costs limit how much new transmission can be built \cite{DOE2024TransmissionRule, NGA2023Transmission}, so where a limited amount of new transmission should be placed becomes a central design question.
Placement matters all the more because the outcome is not always the intuitive one. Braess's paradox, first identified in transportation networks, describes how adding capacity can degrade system performance \cite{braess1968paradoxon}. 
The same effect occurs in oscillator networks and power grids, where a new line can destabilize a previously stable configuration by redistributing flows nonlocally \cite{witthaut2012braess}. 
It has also been demonstrated in realistic grid models, together with a structural criterion for predicting which extensions carry this risk \cite{schafer2022understanding}. 
What remains less clear is how the effect of an added line depends jointly on its placement, the distribution of nodal power, and the arrangement of producers and consumers.

Here we ask how synchronization stability changes when new links, which we refer to as shortcuts, are added to ring networks in which producer and consumer nodes are arranged in clusters.
We use basin stability to quantify how each candidate shortcut affects the robustness of synchronization, and evaluate placements across four scenarios that vary network scale, the degree of decentralization, the heterogeneity of nodal power injections, and the spatial arrangement of producer and consumer clusters.
This comparison identifies placements that enhance synchronization and placements that produce Braess-like degradation, and shows how these outcomes depend on shortcut placement relative to the arrangement of producers and consumers.

\section{Preliminaries}\label{sec:pre}
\subsection{Base Model Structure}
\begin{figure}[!t]
        \centering
    \includegraphics[width=\columnwidth]{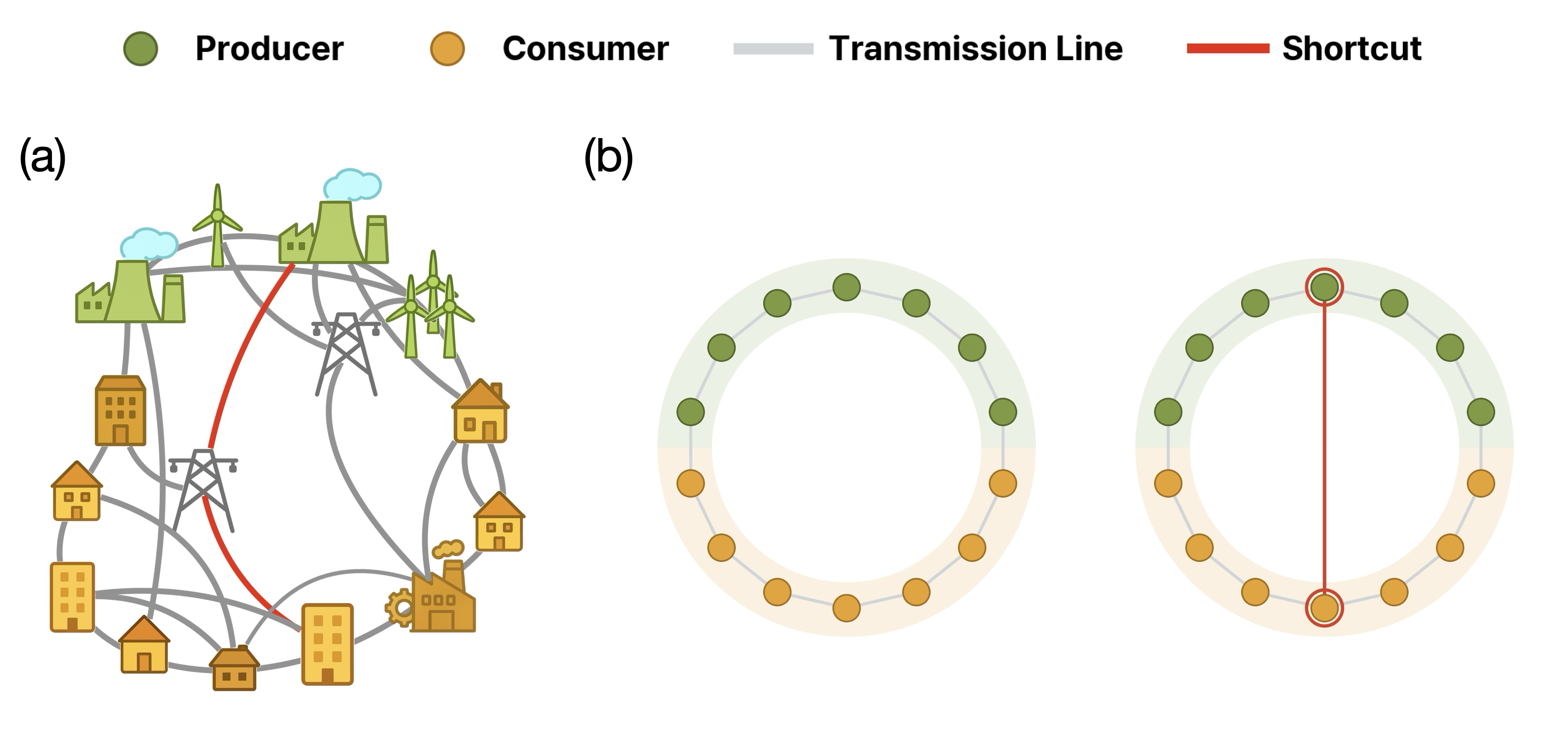}
        \caption{\textbf{Power-grid abstraction and ring-network model.}
        (a) Schematic abstraction of a power grid as a network of nodes and transmission lines, with the red line marking an added long-range connection.
        (b) The ring model used in this study. Producer and consumer nodes occupy two contiguous arcs (left), and a shortcut is added as an extra line between two non-adjacent nodes, here connecting the two cluster centers (right).}
        \label{fig:schem}
    \end{figure}

To examine how topology influences synchronization stability, we use a ring network in which producer and consumer nodes occupy two contiguous arcs (Fig.~\ref{fig:schem}b, left).
The ring serves as a controlled topology rather than a literal transmission grid.
Every node has the same degree and there are no dead ends, so a placement can be moved without also changing the local connectivity of the nodes it joins.
We consider an even number \(N\) of nodes, indexed from \(0\) to \(N-1\) along the ring.
In the base configuration, nodes \(0\) through \(N/2-1\) are producers and nodes \(N/2\) through \(N-1\) are consumers, so the two groups are equal in size and meet at two boundaries.
This arrangement captures, in simplified form, the spatial separation between generation and consumption.
Neighboring nodes are connected by transmission lines of uniform coupling strength.
Total generation matches total consumption, \(\sum_{i=0}^{N-1} P_i = 0\), and transmission losses are neglected.

A shortcut is an additional transmission line placed between two non-adjacent nodes of this ring, assigned the same coupling strength as the existing lines (Fig.~\ref{fig:schem}b, right).
It provides an additional path through which power can be redistributed across the network.
Adding such a line gives a simplified representation of transmission expansion, in which new connections create alternative pathways for power transfer~\cite{DOE2024TransmissionRule, NGA2023Transmission, schafer2022understanding}.
Comparing a network with and without a shortcut shows what one added line does, and because every placement adds one line of the same coupling strength, comparing placements isolates the effect of where that line is attached.

\subsection{Dynamics Modeling of Power Grid Networks}\label{sec:dynamics}

We model the dynamics of this network with second-order coupled-oscillator equations~\cite{filatrella2008analysis, bergen1981structure, nishikawa2015comparative}.
For node \(i\), the phase angle \(\theta_i\) is measured relative to a frame rotating at the rated angular frequency, and \(\omega_i = \dot{\theta}_i\) is the corresponding angular-frequency deviation.
These evolve as
\begin{equation}
\dot{\theta}_i = \omega_i, \qquad
\dot{\omega}_i = P_i - \alpha \omega_i - \sum_{j=0}^{N-1} K_{ij} \sin(\theta_i - \theta_j),
\label{eq:swing}
\end{equation}
where \(P_i\) is the net power injection at node \(i\), positive for producers and negative for consumers, \(\alpha\) is a uniform damping coefficient, and \(K_{ij} = K_{ji}\) is the coupling strength of the line connecting nodes \(i\) and \(j\), which is zero when the two nodes are not connected.
The nodal inertia is normalized to unity.
A synchronized steady state is a fixed point of Eq.~\eqref{eq:swing}, at which \(\omega_i = 0\) for all \(i\) and every node therefore rotates at the rated frequency.
Adding a shortcut sets \(K_{ij}\) for one non-adjacent pair from zero to the value shared by the other lines, leaving all other terms in Eq.~\eqref{eq:swing} unchanged.

\subsection{Basin Stability Analysis}\label{sec:basin}

We quantify stability using basin stability, which measures recovery from finite perturbations and complements linear stability analysis restricted to infinitesimal perturbations~\cite{menck2013basin}.
We use the single-node form of this measure, in which one node is displaced while all remaining nodes are held at the synchronized state~\cite{menck2014dead, schultz2014detours}.
Assuming a uniform perturbation distribution \(\rho_i\) over \(\theta_i \in (-\pi, \pi]\) and \(\omega_i \in [-20, 20]\), the single-node basin stability of node \(i\) is
\begin{equation}
B_i = \int_{\mathcal{B}_i} \rho_i(\theta_i, \omega_i) \, d\theta_i \, d\omega_i ,
\label{eq:bs}
\end{equation}
where \(\mathcal{B}_i\) is the set of perturbed states from which the network returns to synchrony, so that \(B_i \in [0,1]\).

\begin{figure}[!t]
    \centering
    \includegraphics[width=\columnwidth]{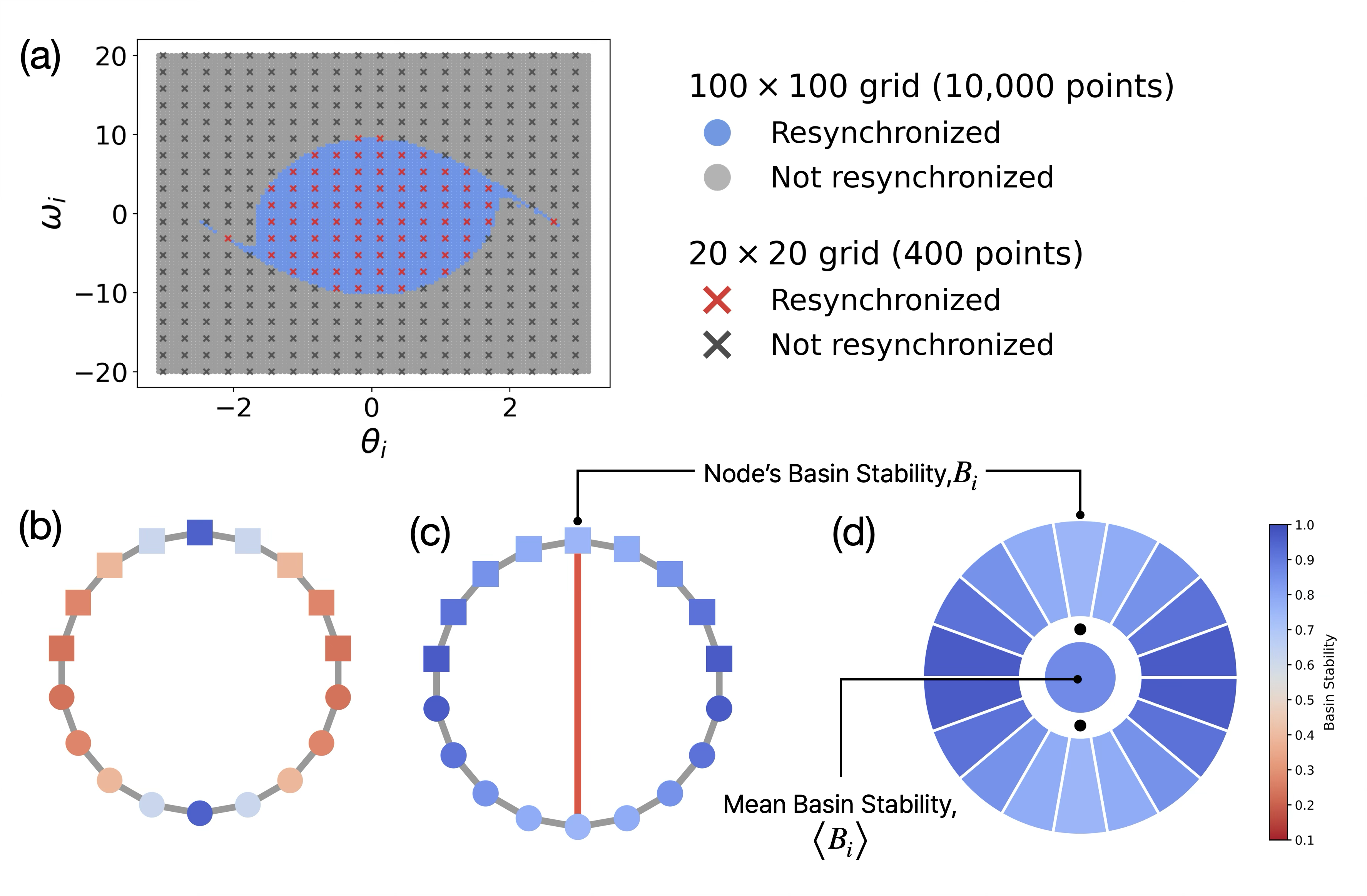}
    \caption{\textbf{Estimating basin stability in the ring-network power-grid model.}
    (a) Perturbation outcomes for a single node in the \(\theta_i\)--\(\omega_i\) plane, comparing the \(20 \times 20\) grid used throughout with a higher-resolution \(100 \times 100\) grid. The fraction of recovering perturbations gives \(B_i = 0.210\) on the coarse grid and \(0.217\) on the fine grid. The node shown lies next to a cluster boundary in a 50-node ring without a shortcut; panels (b)--(d) use an 18-node ring.
    (b) Producer (squares) and consumer (circles) nodes of the ring, each colored by its \(B_i\).
    (c) The same ring with an added shortcut (red line).
    (d) Annular representation of (c): each wedge corresponds to a node at the same angular position and is colored by \(B_i\); the inner circle gives \(\langle B \rangle\), and paired dots between the ring and the inner circle mark the shortcut endpoints.
    The color scale applies to (b)--(d).}
    \label{fig:example}
\end{figure}

Every trajectory is integrated with a fourth-order Runge--Kutta scheme at a fixed step of \(\Delta t = 0.001\) to \(T = 500\), and the network counts as synchronized when \(|\omega_i| < 5 \times 10^{-2}\) for every node at that final time.
A reference synchronized state is obtained by integrating Eq.~\eqref{eq:swing} from \(\theta_i = 0\) and \(\omega_i = 0\) for all \(i\); we then displace a single node from that state and apply the same criterion to classify recovery.
We approximate Eq.~\eqref{eq:bs} on a regular \(20 \times 20\) grid of equally spaced points covering the perturbation range, giving 400 perturbations per node (Fig.~\ref{fig:example}a), so that \(B_i\) is the fraction of these 400 trials that recover synchrony.
To summarize the network-level response we use the mean single-node basin stability,
\begin{equation}
\langle B \rangle = \frac{1}{N} \sum_{i=0}^{N-1} B_i .
\label{eq:meanbs}
\end{equation}
Throughout the study, basin stability is evaluated on a \(20 \times 20\) perturbation grid, which provides a practical balance between computational cost and resolution.
For a representative producer node located at the producer--consumer boundary in an \(N = 50\) ring without a shortcut, the resulting value is \(B_i = 0.210\), compared with \(0.217\) on a \(100 \times 100\) grid of \(10{,}000\) perturbations.
This difference is small relative to the placement-dependent changes examined below.

The damping coefficient is fixed at \(\alpha = 0.1\) throughout.
In Scenarios~1 to~3, a single coupling strength \(K\) is used for all network sizes and shortcut placements within each scenario.
It is chosen as the smallest value for which every configuration in that scenario satisfies the synchronization criterion, and is then held fixed throughout the scenario.
The synchronization thresholds in Scenario~4 differ substantially among configurations, so no single coupling strength places all of them at a comparable margin above their own threshold.
We therefore report two comparisons: one in which each placement is evaluated near its own threshold, located by bisection to a resolution of \(0.05\), and one in which all placements are evaluated at a common \(K\).
The estimated thresholds quantify how the coupling required for synchronization depends on placement, while the fixed-\(K\) comparison isolates placement-dependent differences in basin stability.
The adopted values of \(K\) and the power injections \(P_i\) are given in the corresponding sections.

Node-level basin stability is visualized using annular plots, as illustrated in Fig.~\ref{fig:example}d.

\section{Shortcut Effectiveness Across Network Scenarios}\label{sec:scenarios}

We now examine the effect of shortcut placement across four scenarios that vary network size, power allocation, and producer--consumer arrangement.
Scenario~1 increases the number of nodes while each node keeps a fixed power \(P_i = \pm 1\), so that total generation grows with the network, from 9 at \(N = 18\) to 25 at \(N = 50\).
Scenario~2 increases the number of nodes with total generation held at 25, so that the same aggregate power is distributed over progressively more nodes with smaller-magnitude injections, representing a more decentralized allocation of generation and consumption.
Scenario~3 concentrates a larger share of generation and consumption in a subset of nodes, introducing heterogeneity in nodal power injections.
Scenarios~1 through~3 all keep producers and consumers in two separated clusters; Scenario~4 removes that arrangement by interleaving producer and consumer blocks around the ring.
Together, the four scenarios move from a uniform two-cluster network toward configurations that differ in scale, power allocation, and spatial arrangement, allowing us to distinguish effects that persist across settings from those that depend on a particular configuration.

Basin stability quantifies the probability of recovery but does not by itself characterize the transient trajectory~\cite{lee2026energy}.
We therefore first examine a simple producer--consumer system in the time domain (Section~\ref{chain_vs_ring}), and return to the time domain once more in Scenario~1, where representative shortcut placements are compared on a full two-cluster ring.
The remaining scenarios are then compared primarily through basin stability.

\subsection{Dynamical Illustration of the Shortcut Effect: Chain vs. Ring}\label{chain_vs_ring}

\begin{figure}
    \centering
    \includegraphics[width=.9\columnwidth]{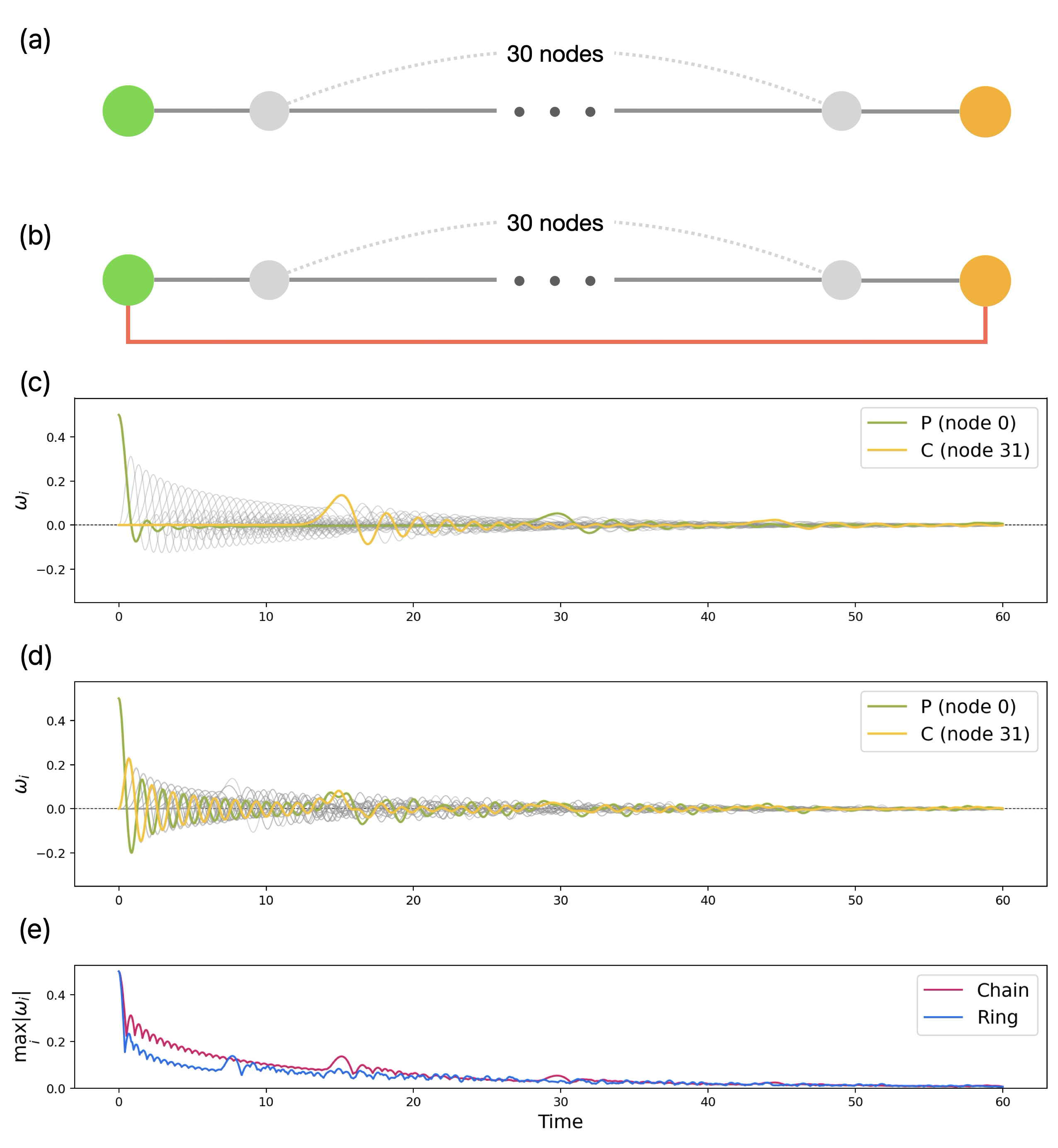}
    \caption{\textbf{Transient response of a producer--consumer chain before and after adding a shortcut.}
    (a) Open chain with a producer \(P\) (green, node 0) and a consumer \(C\) (yellow, node 31) separated by 30 zero-injection nodes (gray).
    (b) Ring formed by adding a direct \(P\)--\(C\) connection (red).
    (c) Angular-frequency deviations \(\omega_i(t)\) in the chain after a perturbation \(\Delta\omega = 0.5\) applied to \(P\) at \(t = 0\). The response of \(C\) is delayed relative to that of \(P\).
    (d) The corresponding response in the ring, where \(C\) responds immediately through the added connection.
    (e) Network-wide maximum angular-frequency deviation \(\max_i |\omega_i(t)|\) for the two topologies. Following the common initial perturbation, the deviation decays more rapidly in the ring over much of the early transient.}
    \label{fig:chain_ring}
\end{figure}

We first illustrate the shortcut effect in a simple producer--consumer chain.
A producer node \(P\) and a consumer node \(C\), with power injections \(P_P = +1\) and \(P_C = -1\), are separated by a chain of 30 zero-injection nodes, so that \(P\) and \(C\) are the only nodes with nonzero power injections (Fig.~\ref{fig:chain_ring}a).
Adding a direct link between \(P\) and \(C\) closes the chain into a ring (Fig.~\ref{fig:chain_ring}b).
This added link directly connects the only producer and consumer in the system.
Both topologies have \(N = 32\) nodes and are integrated with \(K = 5.0\) and \(\alpha = 0.1\).
Starting from the synchronized state, we apply an angular-frequency perturbation \(\Delta\omega = 0.5\) to \(P\) at \(t = 0\).

The two topologies respond to this perturbation in different ways.
In the chain, the disturbance propagates along the intermediate path before reaching the consumer (Fig.~\ref{fig:chain_ring}c).
The consumer stays nearly at rest for the first ten time units, then develops a secondary peak of about \(+0.14\) near \(t \approx 15\), and the transient takes more than 50 time units to decay.
In the ring, the added link connects \(P\) and \(C\) directly, and the consumer responds from the start of the transient (Fig.~\ref{fig:chain_ring}d).
The two nodes then oscillate largely out of phase, so the direct connection produces an immediate counter-response at the consumer rather than one delayed by propagation along the intermediate path.
The network-wide envelope \(\max_i |\omega_i(t)|\) also decays more rapidly in the ring over much of the early transient (Fig.~\ref{fig:chain_ring}e).

This example illustrates the dynamical effect of one particular shortcut, a direct connection between the only producer and consumer.
It does not by itself distinguish the effect of added connectivity from that of shortcut placement.
The following scenarios therefore keep the number of added links fixed while varying their endpoints, so that the role of placement can be examined directly.

\subsection{Scenario 1: Increasing Size with Increasing Total Generation}\label{scenario1}

\begin{figure}
    \centering
    \includegraphics[width=1\columnwidth]{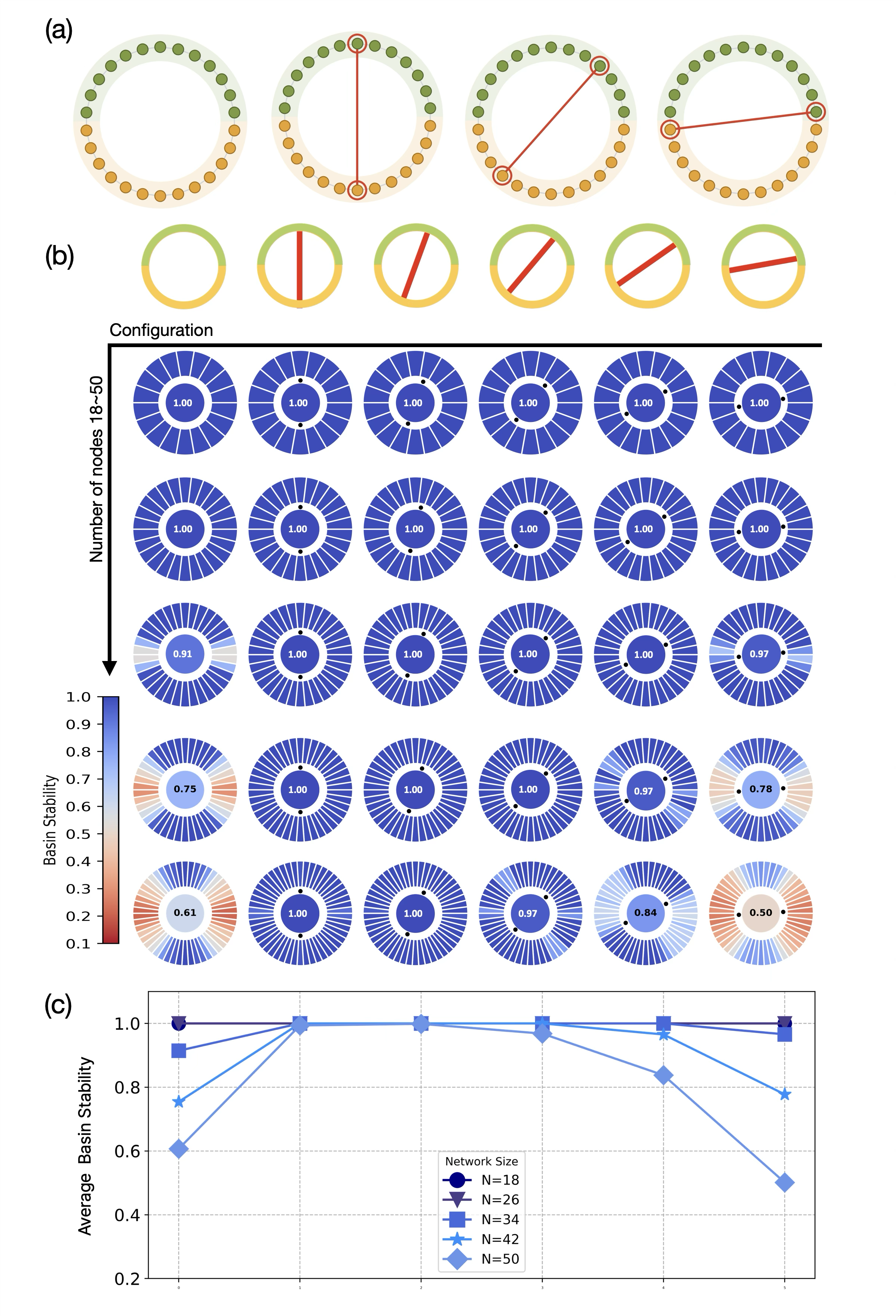}
    \caption{\textbf{Scenario 1: shortcut placement and basin stability under uniform expansion.}
    (a) Two-cluster ring with producers (green) and consumers (orange): the baseline without a shortcut and three of the placements examined, from the center-of-mass (CoM) connection between the two cluster centers to a peripheral connection near the cluster boundaries. Circled nodes mark the shortcut endpoints.
    (b) Basin stability for each placement (columns, ordered from baseline through CoM to peripheral as shown by the icons) and each network size (\(N = 18\)--\(50\), rows). Each annular plot gives the node-level values \(B_i\) as wedges and \(\langle B \rangle\) at the center; paired dots mark the shortcut endpoints.
    (c) \(\langle B \rangle\) for the same placements and sizes, with placements ordered as in (b). The two smallest networks are unaffected by placement, while at \(N = 50\) the same five placements range from \(1.00\) to \(0.50\), the latter below the baseline value of \(0.61\).}
    \label{fig:scen1}
\end{figure}

We first ask how stability responds to network growth, and whether a single added line can offset it.
Circular networks of \(N = 18, 26, 34, 42\), and \(50\) nodes are constructed with equal numbers of producers and consumers, each assigned \(P_i = \pm 1\), so that total generation grows with the network.
The coupling strength is fixed at \(K = 19.5\) across all sizes so that they can be compared directly.
Fig.~\ref{fig:scen1}a shows the two-cluster baseline together with three of the placements examined: a center-of-mass (CoM) shortcut joining the central nodes of the two clusters, an intermediate placement, and a peripheral shortcut joining nodes near the cluster boundaries.
We evaluate five placements spanning this range, shown above each column of Fig.~\ref{fig:scen1}b.

Without a shortcut, the network is fully stable up to \(N = 26\) and then loses stability with each further increase in size, with \(\langle B \rangle\) falling to \(0.91\) at \(N = 34\), \(0.75\) at \(N = 42\), and \(0.61\) at \(N = 50\) (Fig.~\ref{fig:scen1}b, first column; Fig.~\ref{fig:scen1}c).
As the network grows, the ring distance between the two clusters increases while the aggregate power transferred between them also grows, creating conditions under which larger phase angle differences can develop on heavily loaded lines, as discussed in Section~\ref{sec:introduction}.
The node-level values in the same column locate the loss: the low-stability wedges lie at the two boundaries between the producer and consumer arcs, while the nodes at the cluster centers remain stable.
These boundary nodes are the ones farthest along the ring from the center of the opposing cluster.

A CoM shortcut removes this loss entirely.
It yields \(\langle B \rangle = 1.00\) at every size tested, a \(64\%\) gain over the baseline at \(N = 50\), and the boundary nodes that were unstable in the larger baseline networks recover along with it.
Moving the shortcut away from the cluster centers reverses this benefit progressively.
At \(N = 50\), the five placements ordered from CoM to peripheral give \(\langle B \rangle = 1.00\), \(1.00\), \(0.97\), \(0.84\), and \(0.50\), so the most peripheral placement leaves the network less stable than adding no line at all, a decline of about \(18\%\) from \(0.61\).
This is a Braess-like effect: connectivity increases while the probability of recovery decreases.
The node-level pattern of this placement is uneven, with several nodes below their baseline values (Fig.~\ref{fig:scen1}b, bottom right).

This spread depends on network size.
At \(N = 18\) and \(N = 26\) all five placements leave the network at \(\langle B \rangle = 1.00\), so placement is immaterial; the gap between the best and worst placement opens beyond \(N = 34\) and reaches \(0.50\) at \(N = 50\).
Larger networks therefore amplify the cost of a poorly placed line, while the CoM placement maintains \(\langle B \rangle = 1.00\) across all sizes tested.

\begin{figure}
    \centering
    \includegraphics[width=1\columnwidth]{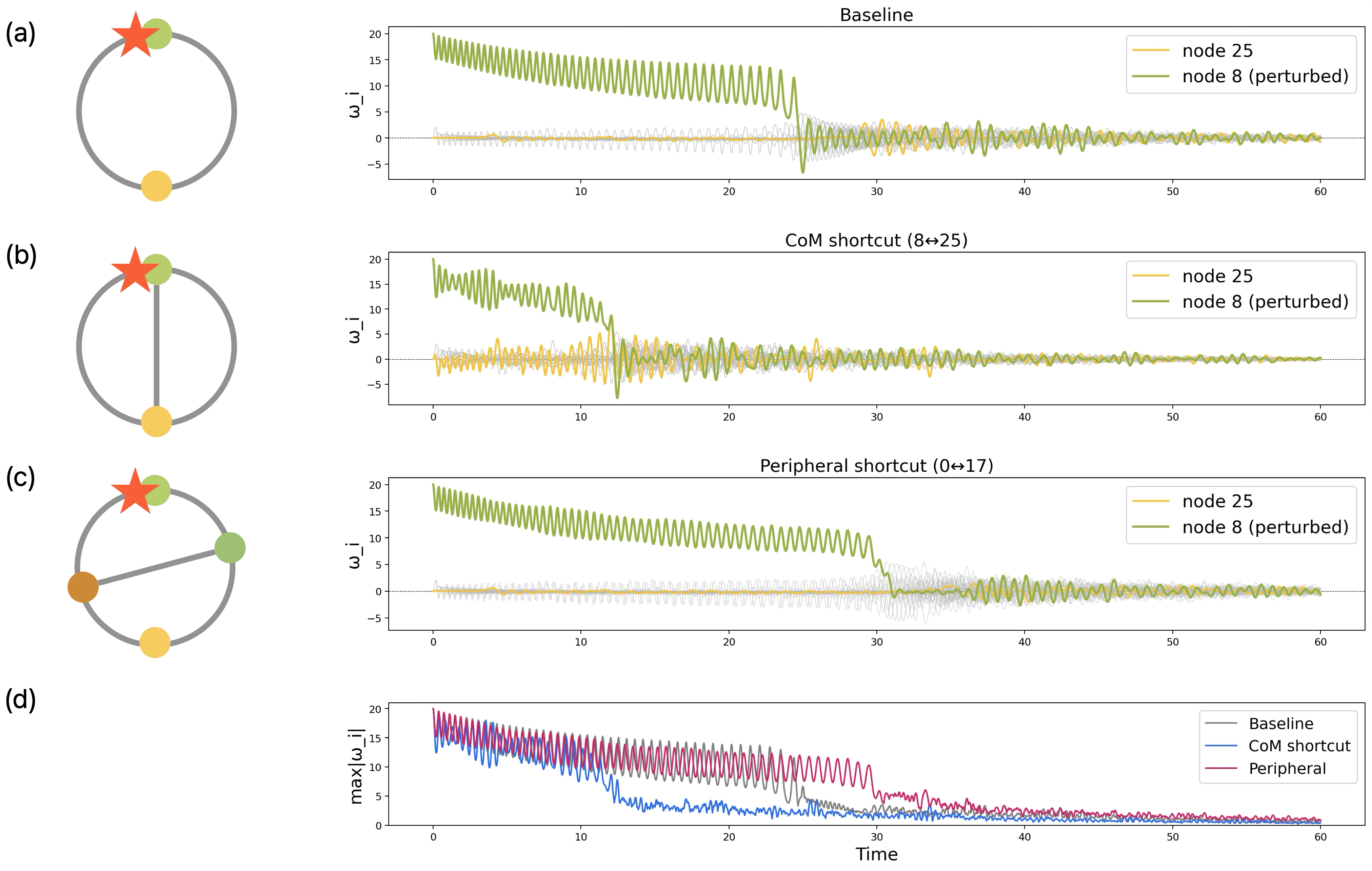}
    \caption{\textbf{Time-domain response of the baseline, CoM, and peripheral configurations at \(N = 34\).}
    An angular-frequency perturbation \(\Delta\omega = 20\) is applied to node 8 at \(t = 0\) in all three configurations. 
    The perturbed node is marked with a star.
    (a) Nodal angular-frequency trajectories for the baseline network without a shortcut.
    (b) Corresponding trajectories with a CoM shortcut connecting the two cluster centers, nodes 8 and 25.
    (c) Corresponding trajectories with a peripheral shortcut connecting nodes 0 and 17 at the cluster boundaries.
    (d) Network-wide maximum angular-frequency deviation \(\max_i |\omega_i(t)|\) for the three configurations.}
    \label{fig:dynamics}
\end{figure}

We next compare the transient responses of the baseline, CoM, and peripheral configurations at \(N = 34\) under an identical angular-frequency perturbation \(\Delta\omega = 20\) applied to node 8, the producer-center node, at \(t = 0\).
A larger perturbation is used here than in Section~\ref{chain_vs_ring} because smaller perturbations recover rapidly in all three configurations and do not clearly separate their transient responses.
The baseline and CoM configurations show a transient contrast similar to that seen in the chain--ring illustration, again in the timing of the response at the opposing cluster.
In the baseline, the perturbed node oscillates at \(15\)--\(20\) for about 25 time units while the disturbance travels around the ring, and the amplitude drops only once it reaches the consumer cluster (Fig.~\ref{fig:dynamics}a); with the CoM shortcut, node 25 at the consumer center responds from the start of the transient and oscillates largely out of phase with node 8, whose oscillation has decayed by \(t \approx 12\) (Fig.~\ref{fig:dynamics}b).
The peripheral shortcut joins nodes 0 and 17, which lie at the two producer--consumer boundaries, so node 25 is not directly connected to the perturbed node and stays nearly inactive through the early transient, as in the baseline.
The rapid damping of node 8 is delayed to \(t \approx 30\), about 5 time units later than in the baseline (Fig.~\ref{fig:dynamics}c), and the network-wide envelope \(\max_i |\omega_i(t)|\) stays above both other configurations for most of the transient (Fig.~\ref{fig:dynamics}d).
For this representative perturbation, the peripheral shortcut therefore produces a slower transient decay, consistent with its lower basin stability.

These results extend the chain--ring illustration to a two-cluster network in which the added line can be placed at different endpoints.
With the number of added links and the coupling strength held fixed, the outcome depends strongly on placement.
A shortcut between the cluster centers raises basin stability and accelerates the representative transient decay, whereas a peripheral shortcut can reduce basin stability below the shortcut-less baseline and slow the transient response.
The next scenario asks whether this placement dependence persists when the same total generation is distributed among more nodes with smaller-magnitude injections.

\subsection{Scenario 2: Increasing Size with Fixed Total Generation}\label{scenario2}
\begin{figure}
        \centering
        \includegraphics[width=1\columnwidth]{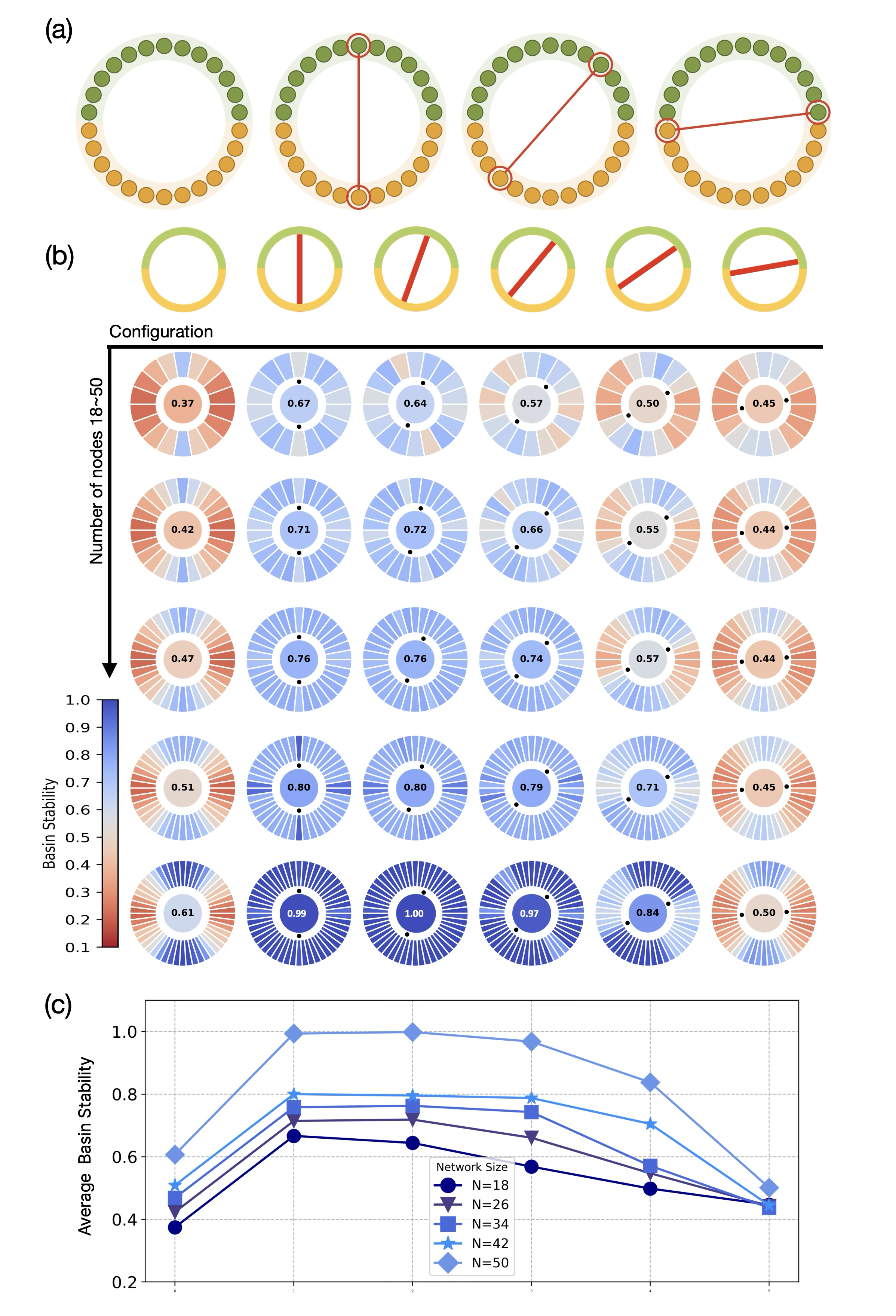}
      \caption{\textbf{Scenario 2: shortcut placement under fixed total generation.}
    (a) Ring networks with fixed total generation \(P_\mathrm{total} = 25\) as the node count increases, so that the nodal injection magnitude \(|P_i| = 25/(N/2)\) decreases with size. Representative shortcut placements range from CoM to peripheral.
    (b) Basin stability for each shortcut placement and network size, in the same format as Fig.~\ref{fig:scen1}b.
    (c) Mean basin stability \(\langle B \rangle\) for the same placements and sizes. The shortcut-less baseline increases from \(0.37\) at \(N = 18\) to \(0.61\) at \(N = 50\), while the peripheral shortcut changes from beneficial at \(N = 18\) (\(0.37 \rightarrow 0.45\)) to detrimental at \(N = 50\) (\(0.61 \rightarrow 0.50\)).}

        \label{fig:scen2}
    \end{figure}

Scenario~2 asks how the shortcut-placement effect changes when the same aggregate generation and consumption are distributed across progressively more nodes.
This corresponds to a shift in generation structure, in which a smaller number of large plants gives way to a larger number of smaller units.
The total generation is fixed at \(P_\mathrm{total} = 25\) while the network size increases from \(N = 18\) to \(N = 50\), so that the magnitude of the nodal power injection scales as \(|P_i| = 25/(N/2)\).
Increasing \(N\) therefore distributes the same aggregate power among more nodes with smaller-magnitude injections, from \(|P_i| \approx 2.8\) at \(N = 18\) to \(|P_i| = 1\) at \(N = 50\).
The coupling strength is fixed at \(K = 19.5\), and the shortcut placements are the same as in Scenario~1 (Fig.~\ref{fig:scen2}a).

The shortcut-less baseline now becomes more stable as the network grows, with \(\langle B \rangle\) increasing from \(0.37\) at \(N = 18\) to \(0.61\) at \(N = 50\) (Fig.~\ref{fig:scen2}c).
This is the opposite of the trend in Scenario~1, where the same increase in node count was accompanied by a growing total generation.
At \(N = 50\) the two scenarios coincide, since \(|P_i| = 1\) there as well.
The two scenarios together indicate that the decline observed in Scenario~1 is not a consequence of node count alone, but depends on how the power is distributed alongside it.

The CoM shortcut remains beneficial at both ends of this transition.
At \(N = 18\) it raises \(\langle B \rangle\) from \(0.37\) to \(0.67\), and at \(N = 50\) from \(0.61\) to \(0.99\).
The absolute increase in \(\langle B \rangle\) is therefore larger at \(N = 50\), where the nodal injections are smaller; the node-level values are also more uniform, with most nodes approaching \(B_i = 1\) (Fig.~\ref{fig:scen2}b).
Across the intermediate sizes, the CoM placement remains among the most stable configurations, indicating that its benefit is not limited to either endpoint of the transition.

The effect of peripheral placement changes in the opposite direction.
At \(N = 18\) the peripheral shortcut raises \(\langle B \rangle\) from \(0.37\) to \(0.45\), whereas at \(N = 50\) it lowers it from \(0.61\) to \(0.50\).
The same placement therefore changes from mildly beneficial to detrimental as the aggregate power is spread over more nodes, producing a Braess-like degradation in the largest network examined.
The intermediate network sizes show the same transition, with the effect of the peripheral shortcut becoming progressively less favorable as \(N\) increases.

These results show that distributing a fixed aggregate power across more nodes reverses the baseline size trend of Scenario~1 without removing the dependence on shortcut placement.
The CoM shortcut remains beneficial at every size, whereas the peripheral shortcut becomes increasingly unfavorable across the configurations examined.
In practical terms, a reinforcement that helps a network with a few large injections can be the wrong choice once the same power is spread over many smaller ones.
The next scenario tests whether this placement dependence persists when the nodal power injections become heterogeneous.

\subsection{Scenario 3: Increasing Size with Heterogeneous Power}\label{scenario3}
Scenario~3 introduces heterogeneity into the nodal power injections while retaining the two-cluster structure of the previous scenarios.
Electrical power grids exhibit spatial heterogeneity; large industrial facilities and utility-scale installations coexist with geographically distributed residential loads.
Starting from the uniform assignment of Scenario~1, one producer is assigned \(20\%\) of the total generation and one consumer \(20\%\) of the total consumption, while the remaining power is distributed uniformly among the other nodes.
We use this setting to examine whether concentrating part of the power at a small number of nodes changes the sensitivity to shortcut placement.
Two variants are compared: in the fixed case the heterogeneous power assignment is held constant as the shortcut is moved, whereas in the adaptive case the high-magnitude injection nodes are repositioned together with the shortcut endpoints.
This comparison separates the geometric position of the shortcut from its position relative to the heterogeneous power distribution, although the two are not varied independently in the adaptive case.
The shortcut-less baseline is identical in the two variants, because the distinction applies only to shortcut-added configurations.

\begin{figure}
        \centering
        \includegraphics[width=1\columnwidth]{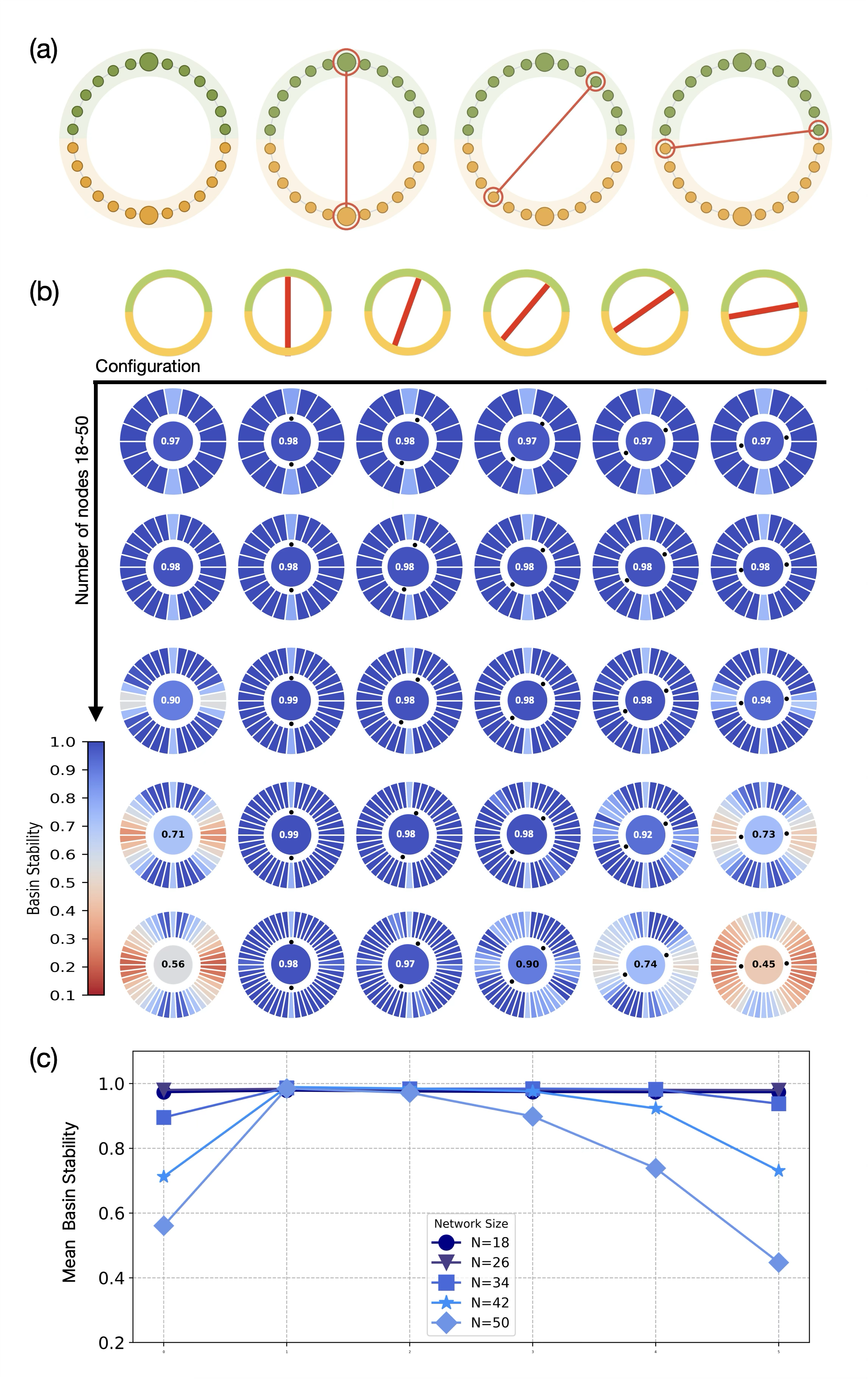}
        \caption{\textbf{Scenario 3a: Heterogeneous power injections with fixed high-magnitude nodes.}
    (a) Ring networks in which one producer and one consumer (enlarged markers) carry 20\% of the total generation and consumption, with the remaining power distributed uniformly.
    These two nodes stay at the two cluster centers regardless of shortcut placement, so that only the CoM placement connects them.
    (b) Basin stability for each shortcut configuration and network size, in the same format as Fig.~\ref{fig:scen1}b.
    (c) Mean basin stability \(\langle B \rangle\) for the same placements and sizes. At \(N = 50\) the CoM shortcut raises \(\langle B \rangle\) from \(0.56\) to \(0.98\), while the most peripheral placement lowers it to \(0.45\).}
    \label{fig:scen3-1}
\end{figure}
\begin{figure}
        \centering
        \includegraphics[width=1\columnwidth]{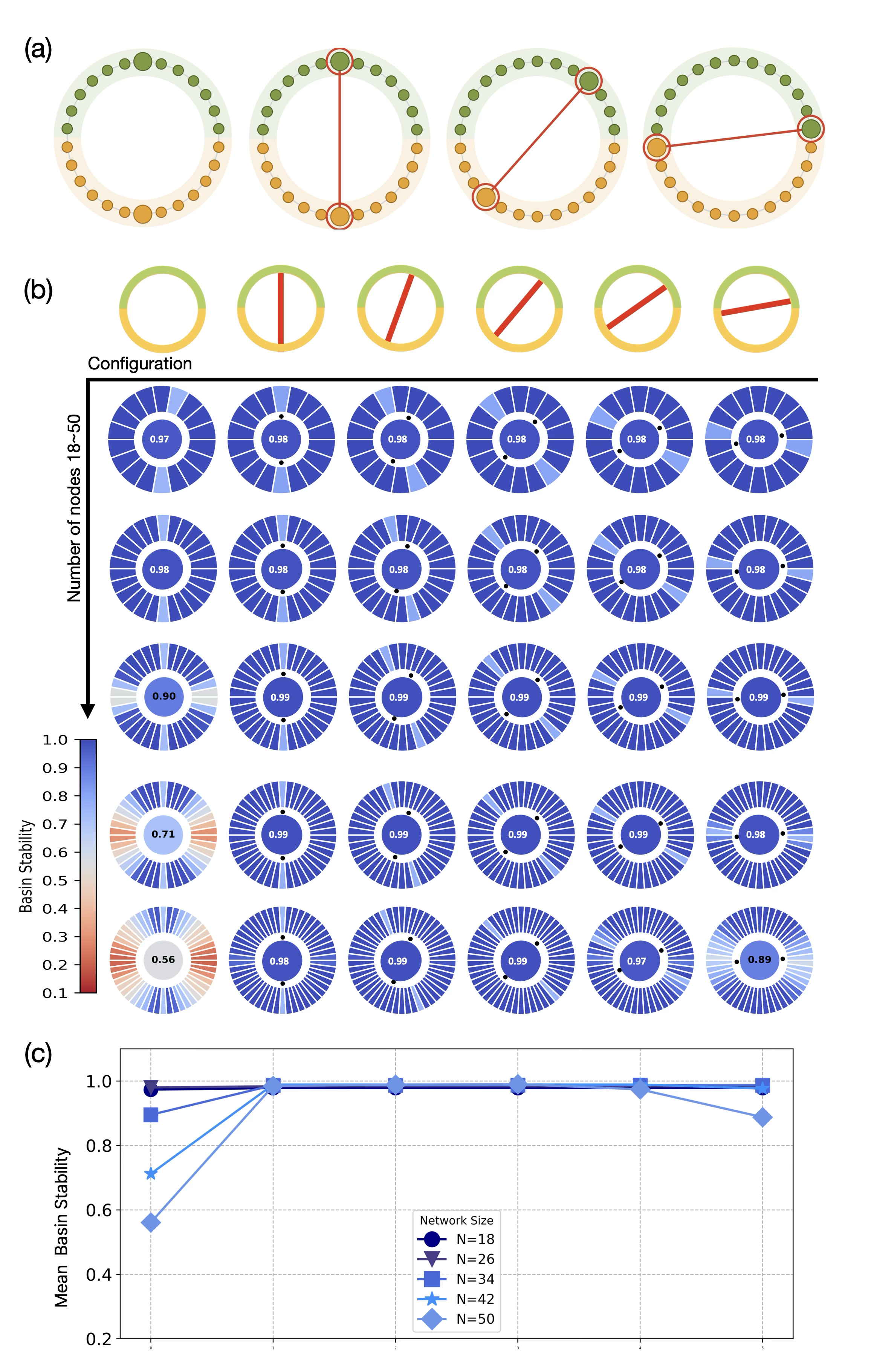}
        \caption{\textbf{Scenario 3b: Heterogeneous power injections with adaptively positioned high-magnitude node.}
    (a) Ring networks with the same distribution of injection magnitudes as in Fig.~\ref{fig:scen3-1}, but with the two high-magnitude nodes repositioned together with the shortcut endpoints, so that every placement connects them.
    (b) Basin stability for each shortcut configuration and network size, in the same format as Fig.~\ref{fig:scen1}b. The baseline column is identical to the fixed-center case (Fig.~\ref{fig:scen3-1}), since the two variants share the same shortcut-less network.
    (c) Mean basin stability \(\langle B \rangle\) for the same placements and sizes. At \(N = 50\) the most peripheral placement gives \(\langle B \rangle = 0.89\), against \(0.45\) for the same placement in the fixed case.}
    \label{fig:scen3-2}
\end{figure}
In the fixed case, shortcut placement has a strong effect on basin stability (Fig.~\ref{fig:scen3-1}).
At \(N = 50\), the shortcut-less baseline gives \(\langle B \rangle = 0.56\), below the \(0.61\) of the uniform network of Scenario~1.
The CoM shortcut raises it to \(0.98\), an increase of \(75\%\) against the \(64\%\) obtained in the uniform case.
Moving the shortcut away from the two high-magnitude nodes progressively reduces the benefit, and the most peripheral placement lowers \(\langle B \rangle\) to \(0.45\), about \(19.6\%\) below the baseline and a steeper relative decline than the \(18\%\) of Scenario~1.
At \(N = 50\), concentrating part of the power in two nodes therefore lowers the baseline and slightly widens the range between the best and worst placements relative to Scenario~1, strengthening the contrast between favorable and unfavorable shortcut positions.

The adaptive case shows a markedly smaller dependence on the geometric position of the shortcut (Fig.~\ref{fig:scen3-2}).
Because the shortcut-less network is identical to that of the fixed case, the \(N = 50\) baseline remains at \(\langle B \rangle = 0.56\).
When the high-magnitude injection nodes are repositioned together with the shortcut endpoints, basin stability remains high across the shortcut placements examined.
For the most peripheral placement at \(N = 50\), \(\langle B \rangle\) reaches \(0.89\), approximately \(59\%\) above the baseline, whereas the same geometric shortcut placement gives \(0.45\), or \(19.6\%\) below the baseline, when the heterogeneous power assignment is held fixed.
Accordingly, the range spanned by the five shortcut placements at \(N = 50\) narrows from \(0.53\) in the fixed case to \(0.09\) in the adaptive case.
This reduced variation indicates that the placement sensitivity observed in the fixed case is closely associated with the shortcut's position relative to the heterogeneous power distribution, rather than with geometric position alone.

Together, the two variants indicate that shortcut effectiveness depends not only on the geometric position of the added line in the ring but also on its position relative to the heterogeneous distribution of nodal power injections.
In practical terms, a line connecting two high-magnitude injection nodes and the same line placed elsewhere in the network are not interchangeable, even when the added capacity is identical.
The next scenario examines whether this placement dependence persists when the producer--consumer arrangement itself is changed.

\subsection{Scenario 4: Topological Variations}\label{scenario4}

Scenario~4 examines shortcut placement when the two-cluster arrangement of the preceding scenarios is replaced by alternating producer and consumer blocks.
In the two-cluster networks, shortcuts connecting the cluster centers consistently performed well, and Scenario~3 showed that their effectiveness also depends on the shortcut position relative to the heterogeneous power distribution.
Here we ask whether the same placement dependence persists when several producer and consumer blocks are distributed around the ring, so that no single shortcut connects the centers of the entire producer and consumer regions.
We therefore vary the position of the shortcut endpoints, the types of blocks they join, and the chord length of the connection.

\begin{figure*}[!t]
    \centering
    \includegraphics[width=6in]{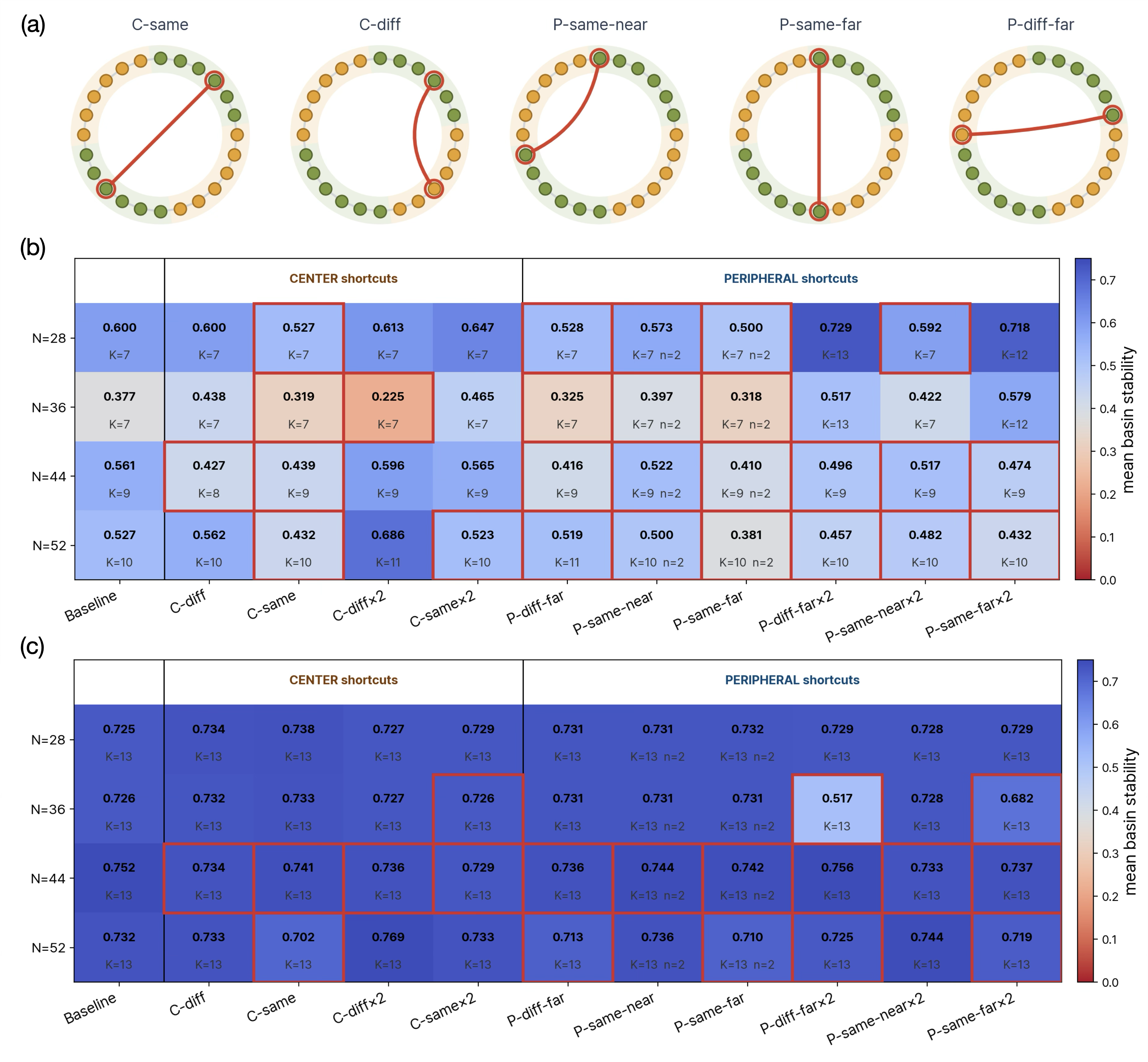}
    \caption{\textbf{Scenario 4: shortcut placement in clustered ring topologies.}
    (a) Shortcut configuration types, classified by the anchoring position of the endpoints (block center, C, or block periphery, P), the types of blocks connected (same or diff), and the chord length of peripheral links (near or far). In doubled variants (\(\times 2\)), the coupling strength assigned to a single added line is divided equally between two symmetrically placed shortcuts, with each shortcut receiving one-half of that coupling strength, so that the total added coupling remains unchanged.
    (b) Mean basin stability of each configuration across network sizes (\(N = 28\)--\(52\)). Each configuration, including the shortcut-less baseline, is evaluated at the smallest integer coupling above its own synchronization threshold, with the adopted \(K\) value indicated in each cell.
    Red frames indicate shortcut configurations whose mean basin stability is below the shortcut-less baseline value of the same size network.
    (c) Mean basin stability of the same configurations evaluated at a common coupling \(K = 13\), the smallest integer value at which all configurations synchronize.}
    \label{fig:scen4}
\end{figure*}

Each shortcut configuration is classified by three geometric attributes: the anchoring position of its endpoints (block centers, C, or block peripheries, P), the types of blocks it joins (same or different, abbreviated diff), and, for peripheral links, the chord length of the connection (near or far) (Fig.~\ref{fig:scen4}a).
A configuration is named by these attributes in that order, so that P-same-far denotes a link between the peripheries of two blocks of the same kind across a long chord.
In this taxonomy, C-diff is the closest analogue of the center shortcut in the alternating-block topology, because it connects the centers of one producer and one consumer block.
Doubled variants are also examined and marked \(\times 2\), in which the coupling strength assigned to one added line is divided equally between two symmetrically placed shortcuts, with each shortcut receiving one-half of that coupling strength, so that the total added coupling remains unchanged.
Networks of \(N = 28\), \(36\), \(44\), and \(52\) contain four alternating blocks of equal size, with two producer and two consumer blocks and uniform nodal injections \(P_i = \pm 1\).

Unlike the preceding scenarios, the alternating-block configurations exhibit substantially different synchronization thresholds depending on shortcut placement.
For each configuration, the threshold is first located by bisection with a resolution of \(0.05\), after which the smallest integer coupling above that threshold is adopted for the basin-stability calculation; these adopted values are shown in each cell of Fig.~\ref{fig:scen4}b.
Because they place the networks at different distances above their synchronization thresholds, we additionally evaluate every configuration at a common coupling \(K = 13\), the smallest integer value at which all configurations synchronize (Fig.~\ref{fig:scen4}c).
Fig.~\ref{fig:scen4}b therefore characterizes each configuration near its own synchronization onset, whereas Fig.~\ref{fig:scen4}c provides a fixed-coupling comparison of shortcut placement.

At their configuration-specific coupling values, none of the single-shortcut configurations has a substantially higher basin stability than the shortcut-less baseline under the same near-onset evaluation protocol, and several have lower values (Fig.~\ref{fig:scen4}b).
The reduction is particularly consistent for the same-kind long-range connections: C-same and P-same-far remain below the ring baseline at every size examined, with changes changes in \(\langle B\rangle\) of approximately \(-0.06\) to \(-0.15\) (red frames in Fig.~\ref{fig:scen4}b).
By contrast, C-diff and P-same-near remain comparatively close to the ring baseline.
The strong enhancement produced by a single center-to-center shortcut in the two-cluster networks is therefore no longer observed when producer and consumer blocks alternate around the ring.

At \(N = 52\), C-diff\(\times 2\), which connects both complementary center pairs, gives \(\langle B \rangle = 0.686\) under the configuration-specific coupling protocol, compared with \(0.527\) for the shortcut-less baseline evaluated near its own synchronization onset (Fig.~\ref{fig:scen4}b, bottom row).
The improvement remains visible at the common coupling, where \(\langle B \rangle\) increases from \(0.732\) to \(0.769\) (Fig.~\ref{fig:scen4}c, bottom row).
Connecting more than one complementary producer--consumer pair can therefore recover part of the benefit that is absent for a single center link.

The peripheral configurations further show that chord length matters even when the endpoint types are unchanged (Fig.~\ref{fig:scen4}b).
P-same-near remains close to the baseline, whereas P-same-far is consistently more detrimental and produces the largest reduction among the single-link configurations, gives \(0.381\) compared with \(0.527\) for the shortcut-less baseline under the configuration-specific protocol.
A related contrast appears among the center connections, where C-diff performs better than C-same.
P-diff-far lies between these cases: connecting different block types reduces, but does not remove, the penalty associated with a long chord.
Taken together, these comparisons indicate that both the types of blocks joined and the chord length of the shortcut contribute to its effect on basin stability.

Splitting the same total added coupling across two long-range shortcuts produces a different effect on the synchronization threshold.
For P-same-far\(\times 2\) and P-diff-far\(\times 2\), the adopted coupling rises to \(K = 12\)--\(13\) at \(N = 28\) and \(36\), whereas the corresponding single-link configurations synchronize at \(K = 7\) (adopted values in Fig.~\ref{fig:scen4}b).
Dividing the added coupling between two diametric chords can therefore increase the coupling required for synchronization even though the total added coupling is unchanged.

The fixed-coupling comparison shows that the higher threshold does not translate into a compensating basin-stability advantage (Fig.~\ref{fig:scen4}c).
At \(K = 13\), the doubled far configurations remain at or below the corresponding ring baseline, and at \(N = 36\) P-diff-far\(\times 2\) lowers \(\langle B \rangle\) from \(0.726\) to \(0.517\).
The same panel also clarifies the particularly low value of C-diff\(\times 2\) at \(N = 36\) in Fig.~\ref{fig:scen4}b: its basin stability increases from \(0.225\) near its configuration-specific threshold to \(0.727\) at the common coupling.
This large change indicates that basin stability in these configurations can depend strongly on the margin above the synchronization threshold.

The \(N = 44\) networks form an exception to several of these trends.
Most shortcut configurations fall below the ring baseline at this size, and the depression is visible in both Fig.~\ref{fig:scen4}b and Fig.~\ref{fig:scen4}c.
Because it persists when \(K\) is held fixed, the effect cannot be attributed solely to the different coupling values used in Fig.~\ref{fig:scen4}b.
Its origin is not resolved here, and we therefore treat \(N = 44\) as a size-specific exception rather than incorporating it into the general trend.

Together, these results show that the shortcut-placement dependence found in the two-cluster networks persists, but takes a different form when producer and consumer blocks alternate around the ring.
No single center-to-center shortcut produces the large enhancement observed in the two-cluster cases, while long same-kind connections are consistently among the least favorable placements.
Both block-type pairing and chord length affect the outcome, and distributing the same added coupling across multiple long-range links can also raise the synchronization threshold.
At the same time, the improvement obtained with C-diff\(\times 2\) at \(N = 52\) shows that complementary center connections can remain beneficial when more than one producer--consumer block pair is connected.
These results therefore limit the simple center connection rule of the earlier scenarios: its effectiveness depends on the spatial organization of producer and consumer regions, and cannot be transferred unchanged to an alternating multi-block topology.

\section{Conclusion}
This study systematically examines how the placement of added transmission links affects synchronization stability in simplified power-grid networks.
Across the two-cluster scenarios, shortcut effectiveness depends strongly on where the added line is placed relative to the producer and consumer clusters.
center-to-center shortcuts did not lower the mean basin stability in the configurations examined and raise it by up to \(75\%\) over the shortcut-less baseline, whereas peripheral shortcuts can reduce it by up to \(28\%\), producing Braess-like degradation.
The direction of the peripheral effect is not fixed, however: under fixed total generation, the corresponding peripheral placement is mildly beneficial at \(N = 18\) but detrimental at \(N = 50\).
Representative time-domain simulations are consistent with the basin-stability results, in that center shortcuts produce faster transient decay for the perturbations examined, whereas peripheral shortcuts lead to slower recovery.

The heterogeneous-power scenario further shows that shortcut effectiveness depends on its position relative to the distribution of nodal power injections.
When the high-magnitude injection nodes remain fixed, moving the shortcut away from them substantially widens the difference between favorable and unfavorable placements.
When these nodes are repositioned together with the shortcut endpoints, the variation among shortcut placements is strongly reduced.
Geometric position alone therefore does not determine the effect of an added line; its position relative to the underlying power distribution is also important.

This placement dependence changes when the two-cluster structure is replaced by alternating producer and consumer blocks.
In this topology, no single center-to-center shortcut produces the large enhancement observed in the two-cluster cases.
Instead, both the types of blocks connected and the chord length of the shortcut affect basin stability.
Long same-kind connections are consistently among the least favorable placements, while connecting more than one complementary producer--consumer center pair can remain beneficial, as the C-diff\(\times 2\) configuration shows at \(N = 52\).
Distributing the same added coupling across multiple long-range links can also increase the coupling required for synchronization.
The simple center to center placement strategy identified in the two-cluster networks therefore cannot be transferred unchanged to alternating multi-block topologies.

The present models are intentionally simplified in order to isolate the effects of network structure and shortcut placement.
Future work should test whether these findings persist in more realistic grid topologies with heterogeneous line parameters, inertia and damping, more detailed power-injection patterns, and realistic control dynamics.
It would also be useful to examine time-varying operating conditions and uncertainty, and to develop optimization- or learning-based methods for identifying effective line additions in larger networks.

Overall, an added transmission link is not interchangeable with the same link placed elsewhere in the network.
What distinguishes them is their position relative to the spatial organization of generation and consumption: in the two-cluster networks examined here, connections between cluster centers consistently outperform peripheral alternatives, whereas alternating-block topologies require consideration of block type, chord length, and the distribution of added coupling.
These findings provide a structural basis for further investigation of transmission-line placement in more realistic power-grid models.
In the two-cluster networks examined here, connections between cluster centers consistently outperform peripheral alternatives, whereas alternating-block topologies require consideration of block type, chord length, and the distribution of added coupling.

\begin{acknowledgments}
This work was supported by the Korea Institute of Energy Technology Evaluation and Planning (KETEP) and the Ministry of Climate, Energy \& Environment (MCEE) of the Republic of Korea (No. RS-2026-25537902).
\end{acknowledgments}

\begin{acknowledgments}
This work was supported by the Korea Institute of Energy Technology Evaluation and Planning (KETEP) and the Ministry of Climate, Energy \& Environment (MCEE) of the Republic of Korea (No. RS-2026-25537902).
\end{acknowledgments}

\section*{AUTHOR DECLARATIONS}

\subsection*{Conflict of Interest}
The authors have no conflicts to disclose.

\subsection*{Author Contributions}

Yunju Choi: Conceptualization; Methodology; Software; Formal analysis; Investigation; Visualization; Writing -- original draft; Writing -- review \& editing.

Daekyung Lee: Methodology; Software; Validation; Writing -- review \& editing.

Heetae Kim: Conceptualization; Methodology; Supervision; Project administration; Funding acquisition; Writing -- review \& editing.

\section*{Data Availability Statement}
The data that support the findings of this study are available from the corresponding author upon reasonable request.

\bibliography{ref}

@article{cohon1980application,
  title={Application of a multiobjective facility location model to power plant siting in a six-state region of the US},
  author={Cohon, Jared L and ReVelle, Charles S and Current, John and Eagles, Thomas and Eberhart, Russell and Church, Richard},
  journal={Computers \& Operations Research},
  volume={7},
  number={1-2},
  pages={107--123},
  year={1980},
  publisher={Elsevier}
}

@article{hoffacker2020local,
  title={Local energy: spatial proximity of energy providers to their power resources},
  author={Hoffacker, Madison K and Hernandez, Rebecca R},
  journal={Frontiers in Sustainability},
  volume={1},
  pages={585110},
  year={2020},
  publisher={Frontiers Media SA}
}

@misc{DOE2024TransmissionRule,
  author       = {{U.S. Department of Energy}},
  title        = {Biden-Harris Administration Announces Final Transmission Permitting Rule},
  howpublished = {DOE Press Release},
  year         = {2024},
  month        = apr,
  day          = {25},
  url          = {https://www.energy.gov/articles/biden-harris-administration-announces-final-transmission-permitting-rule}
}

@misc{NGA2023Transmission,
  author       = {{National Governors Association}},
  title        = {Transmission Siting and Permitting: How Governor Leadership can Advance Projects},
  year         = {2023},
  month        = mar,
  url          = {https://www.nga.org/publications/issue-brief/transmission-siting-and-permitting-how-governor-leadership-can-advance-projects}
}

@article{buldyrev2010catastrophic,
  title={Catastrophic cascade of failures in interdependent networks},
  author={Buldyrev, Sergey V and Parshani, Roni and Paul, Gerald and Stanley, H Eugene and Havlin, Shlomo},
  journal={Nature},
  volume={464},
  number={7291},
  pages={1025--1028},
  year={2010},
  publisher={Nature Publishing Group UK London}
}

@article{kundur2007power,
  title={Power system stability},
  author={Kundur, Prabha},
  journal={Power system stability and control},
  volume={10},
  number={1},
  pages={7--1},
  year={2007}
}

@article{schultz2014detours,
  title={Detours around basin stability in power networks},
  author={Schultz, Paul and Heitzig, Jobst and Kurths, J{\"u}rgen},
  journal={New Journal of Physics},
  volume={16},
  number={12},
  pages={125001},
  year={2014},
  publisher={IOP Publishing}
}

@article{menck2014dead,
  title={How dead ends undermine power grid stability},
  author={Menck, Peter J and Heitzig, Jobst and Kurths, J{\"u}rgen and Joachim Schellnhuber, Hans},
  journal={Nature communications},
  volume={5},
  number={1},
  pages={3969},
  year={2014},
  publisher={Nature Publishing Group UK London}
}

@article{filatrella2008analysis,
  title={Analysis of a power grid using a Kuramoto-like model},
  author={Filatrella, Giovanni and Nielsen, Arne Hejde and Pedersen, Niels Falsig},
  journal={The European Physical Journal B},
  volume={61},
  pages={485--491},
  year={2008},
  publisher={Springer}
}

@article{nishikawa2015comparative,
  title={Comparative analysis of existing models for power-grid synchronization},
  author={Nishikawa, Takashi and Motter, Adilson E},
  journal={New Journal of Physics},
  volume={17},
  number={1},
  pages={015012},
  year={2015},
  publisher={IOP Publishing}
}

@article{dorfler2012synchronization,
  title={Synchronization and transient stability in power networks and nonuniform Kuramoto oscillators},
  author={Dorfler, Florian and Bullo, Francesco},
  journal={SIAM Journal on Control and Optimization},
  volume={50},
  number={3},
  pages={1616--1642},
  year={2012},
  publisher={SIAM}
}

@article{motter2013spontaneous,
  title={Spontaneous synchrony in power-grid networks},
  author={Motter, Adilson E and Myers, Seth A and Anghel, Marian and Nishikawa, Takashi},
  journal={Nature Physics},
  volume={9},
  number={3},
  pages={191--197},
  year={2013},
  publisher={Nature Publishing Group UK London}
}

@article{schafer2022understanding,
  title={Understanding Braess’ paradox in power grids},
  author={Sch{\"a}fer, Benjamin and Pesch, Thiemo and Manik, Debsankha and Gollenstede, Julian and Lin, Guosong and Beck, Hans-Peter and Witthaut, Dirk and Timme, Marc},
  journal={Nature Communications},
  volume={13},
  number={1},
  pages={5396},
  year={2022},
  publisher={Nature Publishing Group UK London}
}

@article{witthaut2012braess,
  title={Braess's paradox in oscillator networks, desynchronization and power outage},
  author={Witthaut, Dirk and Timme, Marc},
  journal={New journal of physics},
  volume={14},
  number={8},
  pages={083036},
  year={2012},
  publisher={IOP Publishing}
}

@article{witthaut2013nonlocal,
  title={Nonlocal failures in complex supply networks by single link additions},
  author={Witthaut, Dirk and Timme, Marc},
  journal={The European Physical Journal B},
  volume={86},
  pages={1--12},
  year={2013},
  publisher={Springer}
}

@article{menck2013basin,
  title={How basin stability complements the linear-stability paradigm},
  author={Menck, Peter J and Heitzig, Jobst and Marwan, Norbert and Kurths, J{\"u}rgen},
  journal={Nature physics},
  volume={9},
  number={2},
  pages={89--92},
  year={2013},
  publisher={Nature Publishing Group UK London}
}

@article{simonsen2008transient,
  title={Transient dynamics increasing network vulnerability to cascading failures},
  author={Simonsen, Ingve and Buzna, Lubos and Peters, Karsten and Bornholdt, Stefan and Helbing, Dirk},
  journal={Physical review letters},
  volume={100},
  number={21},
  pages={218701},
  year={2008},
  publisher={APS}
}

@article{witthaut2016critical,
  title={Critical links and nonlocal rerouting in complex supply networks},
  author={Witthaut, Dirk and Rohden, Martin and Zhang, Xiaozhu and Hallerberg, Sarah and Timme, Marc},
  journal={Physical review letters},
  volume={116},
  number={13},
  pages={138701},
  year={2016},
  publisher={APS}
}

@article{rohden2012self,
  title={Self-organized synchronization in decentralized power grids},
  author={Rohden, Martin and Sorge, Andreas and Timme, Marc and Witthaut, Dirk},
  journal={Physical review letters},
  volume={109},
  number={6},
  pages={064101},
  year={2012},
  publisher={APS}
}

@article{albert2004structural,
  title={Structural vulnerability of the North American power grid},
  author={Albert, R{\'e}ka and Albert, Istv{\'a}n and Nakarado, Gary L},
  journal={Physical review E},
  volume={69},
  number={2},
  pages={025103},
  year={2004},
  publisher={APS}
}

@article{braess1968paradoxon,
    author    = {Braess, Dietrich},
    title     = {{\"U}ber ein Paradoxon aus der Verkehrsplanung},
    journal   = {Unternehmensforschung Operations Research},
    volume    = {12},
    pages     = {258--268},
    year      = {1968},
    doi       = {10.1007/BF01918335}
}

@article{strogatz2000kuramoto,
  title={From Kuramoto to Crawford: exploring the onset of synchronization in populations of coupled oscillators},
  author={Strogatz, Steven H},
  journal={Physica D: Nonlinear Phenomena},
  volume={143},
  number={1-4},
  pages={1--20},
  year={2000},
  publisher={Elsevier}
}

@article{arenas2008synchronization,
  title={Synchronization in complex networks},
  author={Arenas, Alex and D{\'\i}az-Guilera, Albert and Kurths, Jurgen and Moreno, Yamir and Zhou, Changsong},
  journal={Physics reports},
  volume={469},
  number={3},
  pages={93--153},
  year={2008},
  publisher={Elsevier}
}

@article{rodrigues2016kuramoto,
  title={The Kuramoto model in complex networks},
  author={Rodrigues, Francisco A and Peron, Thomas K DM and Ji, Peng and Kurths, J{\"u}rgen},
  journal={Physics Reports},
  volume={610},
  pages={1--98},
  year={2016},
  publisher={Elsevier}
}

@article{tpl2015001,
  title={001-4, Transmission system planning performance requirements},
  author={TPL, NERC Standard},
  journal={Atlanta, Georgia, USA: North American Electric Reliability Corporation (NERC)},
  year={2015}
}

@book{wood2013power,
  title={Power generation, operation, and control},
  author={Wood, Allen J and Wollenberg, Bruce F and Shebl{\'e}, Gerald B},
  year={2013},
  publisher={John wiley \& sons}
}

@article{kim2015community,
  title={Community consistency determines the stability transition window of power-grid nodes},
  author={Kim, Heetae and Lee, Sang Hoon and Holme, Petter},
  journal={New Journal of Physics},
  volume={17},
  number={11},
  pages={113005},
  year={2015},
  publisher={IOP Publishing}
}

@article{varela2001brainweb,
  title={The brainweb: phase synchronization and large-scale integration},
  author={Varela, Francisco and Lachaux, Jean-Philippe and Rodriguez, Eugenio and Martinerie, Jacques},
  journal={Nature reviews neuroscience},
  volume={2},
  number={4},
  pages={229--239},
  year={2001},
  publisher={Nature Publishing Group UK London}
}

@article{yamaguchi2003synchronization,
  title={Synchronization of cellular clocks in the suprachiasmatic nucleus},
  author={Yamaguchi, Shun and Isejima, Hiromi and Matsuo, Takuya and Okura, Ryusuke and Yagita, Kazuhiro and Kobayashi, Masaki and Okamura, Hitoshi},
  journal={Science},
  volume={302},
  number={5649},
  pages={1408--1412},
  year={2003},
  publisher={American Association for the Advancement of Science}
}

@article{lee2026energy,
  title={Energy-barrier characterization of secured basins in coupled oscillators with inertia},
  author={Lee, Jaiyong and Lee, Daekyung and Son, Seung-Woo and Lee, Sang Hoon and Lee, Mi Jin and Kim, Heetae},
  journal={arXiv preprint arXiv:2610.11147},
  year={2026}
}

@article{bergen1981structure,
  title={A structure preserving model for power system stability analysis},
  author={Bergen, Arthur R and Hill, David J},
  journal={IEEE transactions on power apparatus and systems},
  number={1},
  pages={25--35},
  year={1981},
  publisher={IEEE}
}

@article{witthaut2022collective,
  title={Collective nonlinear dynamics and self-organization in decentralized power grids},
  author={Witthaut, Dirk and Hellmann, Frank and Kurths, J{\"u}rgen and Kettemann, Stefan and Meyer-Ortmanns, Hildegard and Timme, Marc},
  journal={Reviews of modern physics},
  volume={94},
  number={1},
  pages={015005},
  year={2022},
  publisher={APS}
}

@article{taylor2009dynamical,
  title={Dynamical quorum sensing and synchronization in large populations of chemical oscillators},
  author={Taylor, Annette F and Tinsley, Mark R and Wang, Fang and Huang, Zhaoyang and Showalter, Kenneth},
  journal={science},
  volume={323},
  number={5914},
  pages={614--617},
  year={2009},
  publisher={American Association for the Advancement of Science}
}

@article{gonze2005spontaneous,
  title={Spontaneous synchronization of coupled circadian oscillators},
  author={Gonze, Didier and Bernard, Samuel and Waltermann, Christian and Kramer, Achim and Herzel, Hanspeter},
  journal={Biophysical journal},
  volume={89},
  number={1},
  pages={120--129},
  year={2005},
  publisher={Elsevier}
}

@article{kiss2002emerging,
  title={Emerging coherence in a population of chemical oscillators},
  author={Kiss, Istv{\'a}n Z and Zhai, Yumei and Hudson, John L},
  journal={Science},
  volume={296},
  number={5573},
  pages={1676--1678},
  year={2002},
  publisher={American Association for the Advancement of Science}
}

@article{winfree1967biological,
  title={Biological rhythms and the behavior of populations of coupled oscillators},
  author={Winfree, Arthur T},
  journal={Journal of theoretical biology},
  volume={16},
  number={1},
  pages={15--42},
  year={1967},
  publisher={Elsevier}
}

@article{breakspear2010generative,
  title={Generative models of cortical oscillations: neurobiological implications of the Kuramoto model},
  author={Breakspear, Michael and Heitmann, Stewart and Daffertshofer, Andreas},
  journal={Frontiers in human neuroscience},
  volume={4},
  pages={190},
  year={2010},
  publisher={Frontiers Research Foundation}
}

@article{buzsaki2004neuronal,
  title={Neuronal oscillations in cortical networks},
  author={Buzsaki, Gyorgy and Draguhn, Andreas},
  journal={science},
  volume={304},
  number={5679},
  pages={1926--1929},
  year={2004},
  publisher={American Association for the Advancement of Science}
}
\end{document}